\documentclass[pra,twocolumn,showpacs,superscriptaddress,floatfix]{revtex4-2}
\usepackage{graphicx}
\usepackage{epstopdf} 
\usepackage{amsmath}
\usepackage{epsfig}
\usepackage{latexsym}
\usepackage{amsfonts}
\usepackage{graphics}
\usepackage{bm}
\usepackage{braket}
\usepackage{dsfont}
\usepackage{soul}
\usepackage{ulem}
\usepackage{bbm}

\usepackage{mathtools}

\usepackage{helvet}

\graphicspath{ {./Graphics/} }
\begin{document}

\date{\today}
\author{J. Mumford}
\affiliation{Department of Physics and Astronomy, McMaster University, 1280 Main Street West, Hamilton, Ontario, Canada L8S 4M1}

\title{Two-particle topological Thouless spin pump}

\begin{abstract}
We show that two particles interacting via spin exchange exhibit topological features found in one-dimensional single particle lattice models.  This is accomplished by absorbing all of the spatial degrees of freedom of the lattices into the spin degrees of freedom of the two particles.  Comparing the spin system with the Su-Schrieffer-Heeger model, we show the existence of topologically protected edge spin states and establish the bulk-edge correspondence.  Modifying the spin system with a chiral symmetry breaking term results in it resembling the Rice-Mele model and can therefore act as a Thouless spin pump of one of the particles when periodically and adiabatically driven.  By using the spin states as a synthetic spatial dimension, we show two particles are enough to simulate well known topological properties in condensed matter physics.
\end{abstract}

\pacs{}
\maketitle

\section{\label{Sec:Intro}Introduction}

Accurately simulating the degrees of freedom and interactions of large quantum systems is difficult or impossible for classical computers due to their complexity.  This is because for most quantum systems the amount of memory required to simulate evolution or even represent a single state of the system increases exponentially with the number of particles.  In recent years, the motivation to skirt these issues has led to other quantum systems being used as quantum simulators \cite{buluta09,schaetz13,georgescu14} because they have quantum rules built into them which do not need to be implemented manually like in classical computers.  Some examples are trapped ions \cite{porras04,kim10,blatt12} and ultracold atoms in optical lattices \cite{lewenstein07,lewenstein12,jaksch05,bloch08} because they are relatively easy to control allowing for systematic study.   The high degree of control has lead to a wide array of systems which can be simulated in fields as far ranging as statistical mechanics \cite{greiner02} and astrophysics \cite{giovanazzi05}.  However, the main direction in which quantum simulations are directed is in solid state physics with a focus on simulating the motion of charged particles in materials \cite{dalibard11,goldman14a,goldman14b,eckardt15,bukov15,creffield16,aidelsburger11}.  Experiments along these lines vary from simulating well known phenomena such as the Meissner effect in an optical ladder lattice \cite{atala14} to creating new exotic states of matter \cite{oka19}.  They have also made it possible to achieve magnetic field strengths previously unattainable in real materials which has led to the observation of the Harper-Hofstadter model \cite{harper55,aidelsburger13,miyake13}.

One process which is of particular interest is the Thouless charge pump \cite{thouless83} which is motion produced from the adiabatic cyclic driving of the lattice the charges are in.  The uniqueness of the Thouless pump comes from the fact that it does not displace charges continuously like electric or magnetic fields, but instead displaces them in clumps and is therefore quantized due to the topology of the pump cycle.  Simulations of the Thouless pump have been successfully realized in, among others systems, ultracold atoms in optical lattices \cite{lohse16,nakajima16,lu16} and in a photonic system with resonator arrays \cite{tang16}.  Thouless \textit{spin} pumps have also been implemented using ultracold atoms in optical lattices \cite{schweizer16} where particles of opposite spin move in opposite direction, so there is no charge transport.

In this work, we simulate the Thouless spin pump by completely internalizing the spatial degrees of freedom of the lattice into spin states.  The system consists of just two particles: one spin-$S$ particle where $S>1$ and one spin-1/2 particle.  We show that two spins interacting via spin exchange is equivalent to an inhomogeneous version of the Su-Schrieffer-Heeger (SSH) model \cite{su79} which is a one-dimensional (1D) lattice model with a topologically nontrivial phase.  The spin system shares the topological features of the SSH model which manifest in the form of protected extreme spin states.  By modifying the spin system, we show that it is equivalent to the inhomogeneous Rice-Mele (RM) model \cite{rice82} which is another 1D lattice model well known for being an example of a Thouless charge pump.  The spin system also shares the pumping behavior in the form of a spin pump of the spin-$S$ particle.  The mimicking of the spatial degrees of freedom with spin states is an example of the use of synthetic dimensions \cite{boada12,celi14,ozawa19b}.  Synthetic dimensions are a useful tool in constructing quantum simulators because they provide versitility in their construction and open up new possibilities of their applicability.  They have been employed in experiments using single particle states such as spin \cite{mancini15,stuhl15,anisimovas16}, momentum \cite{an17,meier16,xie19}, harmonic oscillator \cite{price17}, and rotational \cite{flob15} states of particles.  Recently, many-body cavity modes of light have also been used to simulate both one-dimensional (1D) and two-dimensional (2D) lattices \cite{deng22} and many-body states of matter have been suggested in constructing synthetic lattices \cite{mumford22a,mumford22b}.  The goal of this paper is to use synthetic dimensions to highlight a different way to simulate quantized motion - one without real-space lattices.

\section{\label{model}Model}

\begin{figure*}[t]
\centering
\includegraphics[scale=0.6]{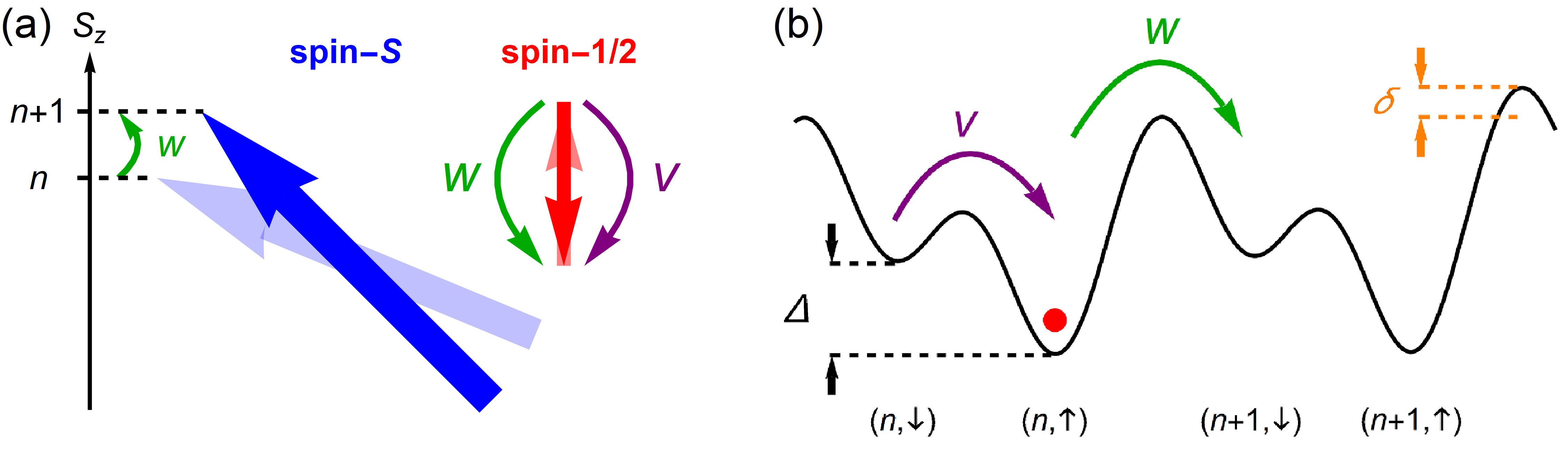}
\caption{Two equivalent nontrivial topological systems.  (a) Giant spin-$S$ particle interacting via spin exchange with a spin-1/2 particle.  The spin-1/2 particle can make transitions on its own ($v$, purple), however, the spin-$S$ particle can only gain or lose one quantum of spin by exchanging it with the spin-1/2 particle ($w$, green).  (b) Single particle in a one-dimensional (1D) lattice.  The particle can tunnel through the small barrier ($v$, purple) or the large barrier ($w$, green).  There is a constant onsite energy offset between adjacent sites labeled $\Delta$ and a tunneling offset that changes along the lattice labeled $\delta$.   The energy offset $\Delta$ acts as a magnetic field applied to the spin-1/2 particle only (not shown) and biases it toward the up or down state depending on its sign.  The tunneling offset $\delta$ is applied to the lattice to match the spin-$S$ transitions which are not homogeneous when $S>1$.}
\label{fig:SL}
\end{figure*}

The system we will be investigating consists of a giant spin-$S$ particle interacting via spin exchange with a spin-1/2 particle which can also flip its spin on its own due to an external field.  The Hamiltonian of the system is 

\begin{equation}
\hat{H}_1 = - w \left (\hat{S}_+ \hat{\sigma}_- + \hat{S}_- \hat{\sigma}_+ \right ) -S v \hat{\sigma}_x
\label{eq:ham}
\end{equation}
where the $S$ operators belong to the spin-$S$ particle and the Pauli spin matrices belong to the spin-1/2 particle.  The factors $w$ and $v$ are the energies for the spin exchange and the spin-1/2 spin-flip processes, respectively.  For $S \gg 1$ we can perform a mean field approximation of the spin-$S$ operators by replacing them with their spin-coherent expectation values $\langle \hat{S}_\pm \rangle =  \sqrt{S^2-n^2} e^{\pm i \phi}$ where $\phi$ is the azimuthal angle on a Bloch sphere of radius $S$ and $n$ is the projection of the spin onto the $S_z$ axis of the Bloch sphere, so $\hat{S}_z \vert n \rangle = n \vert n\rangle$.  The Hamiltonian becomes

\begin{equation}
H_1(n, \phi) = - w \sqrt{S^2 - n^2} \left ( e^{i \phi} \hat{\sigma}_- + e^{-i \phi} \hat{\sigma}_+ \right ) -Sv\hat{\sigma}_x  \, .
\label{eq:ham2}
\end{equation}

The Su-Schrieffer-Heeger (SSH) model Bloch Hamiltonian is

\begin{equation}
H(k) =  - w  \left ( e^{i k} \hat{\sigma}_- + e^{-i k} \hat{\sigma}_+ \right ) - v\hat{\sigma}_x \, ,
\end{equation}
which describes a single particle in a 1D lattice with two sites per unit cell.  Here, the parameters $w$ and $v$ represent the tunneling energies between unit cells and within unit cells, respectively. Due to the SSH model having discrete translational invariance, the quasimomentum $k$ is a good quantum number. Comparing the two Hamiltonians, we see that $\phi$ is similar to $k$, so it can be thought of as the quasimomentum of a spin-state lattice.  Further comparisons show that the spin exchange process can be thought of as tunneling between adjacent unit cells and the spin-1/2 spin-flip process can be thought of as tunneling between the sites within a unit cell.  This means that $n$ is similar to the unit cell number in the SSH model and the two states of the spin-1/2 particle label the two sites in each unit cell.  The obvious difference between the two systems is the square root factor in Eq.\ \eqref{eq:ham2}.  The source of the square root inhomogeneity comes from the boson stimulation factors that arise when making transitions between adjacent spin-$S$ states

\begin{equation}
\hat{S}_\pm \vert n \rangle = \sqrt{\left (S \mp n \right ) \left (S \pm n +1  \right )}\, \vert n \pm 1\rangle \, .
\label{eq:SR}
\end{equation}
Therefore, the two models are equivalent when the SSH model has broken translational invariance and the tunneling is unit cell dependent.  Figure \ref{fig:SL} shows diagrams of the two systems as well as the processes which correspond between them.  Although we cannot use "unit cell" in the strict sense of the term due to the inhomogeneity, we will continue to use it for convenience when making comparisons between the two systems. 

\begin{figure*}[t]
\centering
\includegraphics[scale=0.6]{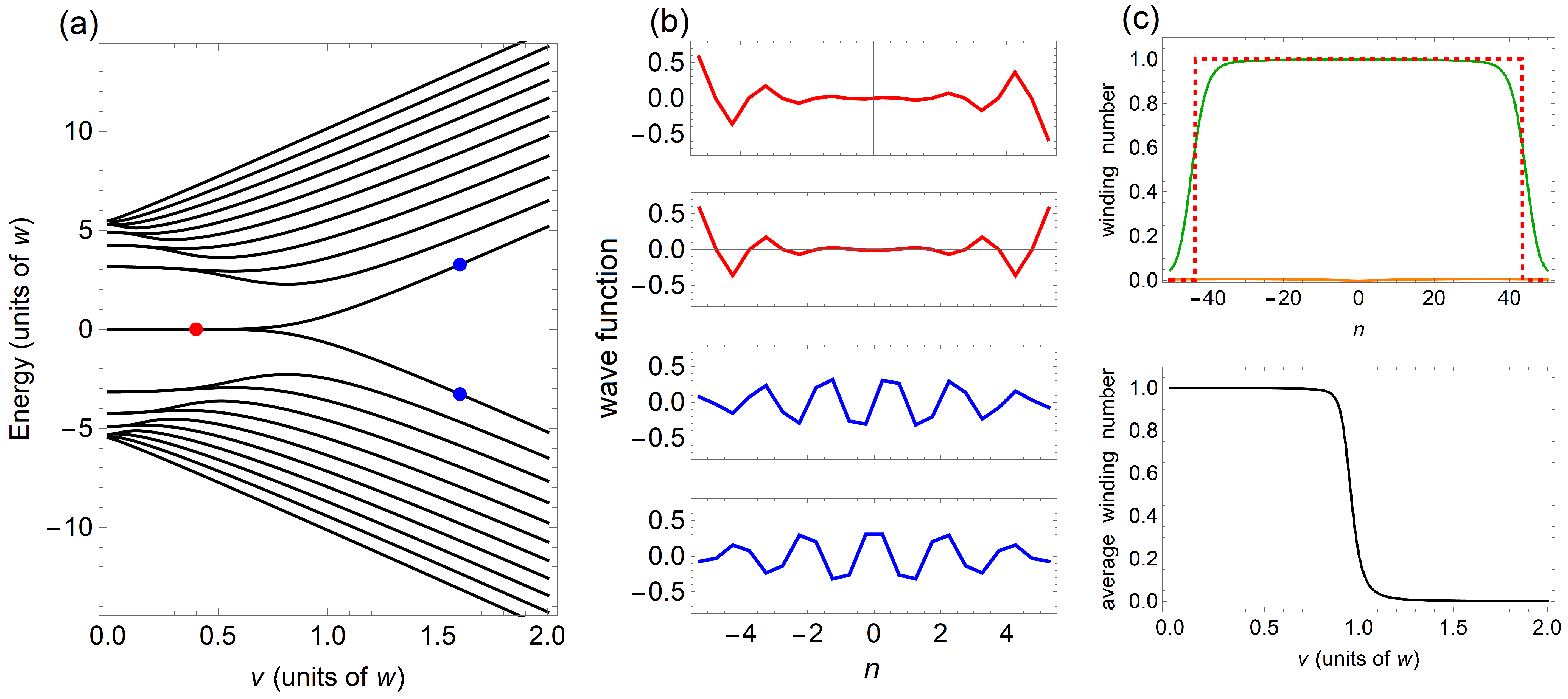}
\caption{Energies, wave functions and winding numbers. (a) Spectrum of the spin system in Eq.\ \eqref{eq:ham} as a function of the spin-1/2 spin-flip energy $v$ for $S = 5$.  The spectrum features doubly degenerate $E = 0$ edge states up until $v \approx w$ where they begin to significantly diverge.  (b) Top and middle-top panels show the wave functions of the $E = 0$, $v = 0.4 w$ [red dot in (a)] edge states while the middle-bottom and bottom panels show the edge states have been destroyed for $v = 1.6w$ [blue dots in (a)].  (c) Top panel shows the localized winding number on either side of the transition point calculated from Eq. \eqref{eq:WN3}: $v = 0.5w$ (green) and $v = 1.5w$ (orange).  The red dashed curve is the mean field winding number $\nu_\mathrm{MF}(n)$ for $v = 0.5w$. Bottom panel shows the average winding number as a function of $v$ for $S = 50$ which has a sudden drop around $v = w $ signifying a topological transition.  The average is taken over the range $-10 \leq n \leq 10$.}
\label{fig:BP}
\end{figure*}

The SSH model is significant because it can display non-trivial topological properties in the form of stable zero energy edge states.  The topological phase of the system can be quantified in terms of the winding number 

\begin{eqnarray}
\nu &=& \frac{i}{2\pi} \int_{0}^{2\pi} dk \, h(k)^{-1} \partial_k h(k) \nonumber \\
&=& \frac{1}{2} \left [1+\mathrm{sgn}(w-v) \right ]
\label{eq:WN1}
\end{eqnarray}
where $\gamma = i\int_{0}^{2\pi} dk \, h(k)^{-1} \partial_k h(k)$ is the Berry phase over the Brillouin zone and $h(k) = v+ w e^{-ik}$.  Here, $w>v$ represents the topological phase with $\nu = 1$ and one edge state at each edge of the lattice whereas $w<v$ represents the trivial phase with $\nu = 0$ and no edge states.  The winding number assumes periodic boundary conditions, so it is a bulk quantity.  The one-to-one correspondence between the winding number and the number of edge states is a fundamental property of 1D topological systems.   In the following section we will see what topological properties of the SSH model are shared by the spin system as well as the effects of the square root factors on those properties.

\section{\label{results} Results}
\subsection{Topological transition}

We begin by looking at the spectrum of the spin model in Fig.\ \ref{fig:BP}(a) for different values of the spin-1/2 transition energy, $v$.  Like the SSH spectrum, it is symmetric about the $E = 0$ axis because both models preserve chiral symmetry $\hat{\mathcal{C}}\hat{H}_1 \hat{\mathcal{C}}^{-1} = - \hat{H}_1$, with $\hat{\mathcal{C}} = \hat{\sigma}_z$.  This symmetry means that for every energy eigenstate $\vert E \rangle$, there is a partner state $\hat{\sigma}_z \vert E \rangle = \vert -E \rangle$.  Differences between the two models are most pronounced at $v = 0$ where in the SSH model there is $2S$-fold degeneracy at  $E_\pm = \pm w$ because the lattice is broken up into $2S$ identical disconnected sublattices of two sites each.  A similar thing happens in the spin model when $v=0$: for a giant spin-$S$ particle the system breaks-up into $2S$ pairs of states $\vert n, \uparrow \rangle$ and $\vert n+1, \downarrow \rangle$ where $\hat{\sigma}_z \vert \uparrow (\downarrow) \rangle = + (-)\vert \uparrow (\downarrow) \rangle$.  Even and odd combinations of these states form the basis and for large $S$ have energies $E_\pm(n) \approx \pm w \sqrt{S^2-n^2}$, so the large degeneracy found in the SSH model is not present.  For $v<w$ the states with $E \neq 0$ show signs of near double degeneracy which comes from the fact that $E_\pm(n) \approx E_\pm(-n)$.

For $v = 0$, the zero energy states are the edge spin states $\vert S, \uparrow \rangle$ and $\vert -S, \downarrow \rangle$.  As $v$ increases, the zero energy states continue to be localized at the edges because their amplitudes within the bulk are suppressed due to the large gap above and below $E = 0$.  The top and middle-top panels of Fig.\ \ref{fig:BP}(b) (red) show the doubly degenerate zero energy wave functions for $v = 0.4$ which are even and odd combinations of states localized at the edges.  For comparison, the bottom and middle-bottom panels (blue) show the same states for $v = 1.6$ where now they mainly occupy the bulk spin states and have $E \neq 0$.  Focusing back on (a), we see that at $v \approx w$ the zero energy states diverge and their gaps between neighboring states shrink signaling the delocalization of these states from the edge.  

Whether this transition has topological origins like the similar one in the SSH model requires the establishment of the bulk-edge correspondence. For that, we need to calculate the winding number and determine if it corresponds to the number of edge states.  To start, the mean field winding number is calculated by using Eq.\ \eqref{eq:WN1} by making the substitutions $h(k) \to h(n, \phi) = Sv + w\sqrt{S^2-n^2}e^{-i\phi}$ and $\partial_k \to \partial_\phi$ and integrating over $\phi$ giving

\begin{equation}
\nu_{\mathrm{MF}}(n) = \frac{1}{2} \left [1+ \mathrm{sgn} \left ( w \sqrt{S^2 - n^2} - Sv \right )\right] \, .
\label{eq:MFWN}
\end{equation}
The winding number for the SSH model in Eq.\ \eqref{eq:WN1} which describes the entire bulk is now spin (site) dependent due to the inhomogeneity.  A numerical calculation from the full quantum model in Eq.\ \eqref{eq:ham} is also performed using the winding number operator \cite{song14,shem14}

\begin{equation}
\hat{\nu} = 4 \hat{P}_{+-}^{-1} \left [ \hat{S}_z, \hat{P}_{+-} \right ]
\label{eq:WN2}
\end{equation}
where $\hat{P}_{+-} = \hat{P}_{-+}^{-1} = \hat{\sigma}_+ \hat{P} \hat{\sigma}_-$ is the off-diagonal spin-1/2 component of the projection operator, $\hat{P}$, of states with $E < 0$.  Comparing $\hat{\nu}$ with Eq.\ \eqref{eq:WN1} we have replaced $h(\phi)$ with $\hat{P}_{\uparrow\downarrow}$ and made use of the fact that $\phi$ and $n$ are conjugate variables, so $\partial_\phi$ is the same as $-i [ \hat{S}_z, \,  ]$.  The integral over $\phi$ can be replaced with a trace over the bulk spin states, however, we localize it to a unit cell to compare it to Eq.\ \eqref{eq:MFWN}

\begin{equation}
\nu (n) = \mathrm{Tr}_\sigma  [\langle n \vert \hat{\nu} \vert n \rangle ]
\label{eq:WN3}
\end{equation}
where $\mathrm{Tr}_\sigma$ is the trace over the spin-1/2 degrees of freedom.  Equation \eqref{eq:WN3} is plotted in the top panel of Fig.\ \ref{fig:BP}(c) for $v > w$ (orange) and $v < w$ (green).  There is a clear difference between the two cases where the bulk has a winding number of unity when $v =0.5w$ and of zero when $v=1.5w$.  The mean field winding number is also plotted as a red dashed curve for $v =0.5w$ where both it and the quantum winding number show clear signs of being bulk quantities as they drop to zero near the edges. The mean field winding number also provides a good approximation of where the drop takes place.  To get a complete picture of the transition, we plot the bulk average winding number in the bottom panel of (c).  The average is taken over the range $-10 \leq n \leq 10$ where $S = 50$, so it is comfortably away from the edges.  A sudden drop from unity to zero is shown at $v \approx w$ agreeing with the vanishing of edge states shown in Fig.\ \ref{fig:BP}(a) and (b).  This establishes the bulk-edge correspondence in the spin system and shows that at $v = w$ there is a topological transition.

\subsection{Spin pump}

\begin{figure}[t]
\centering
\includegraphics[width=\columnwidth]{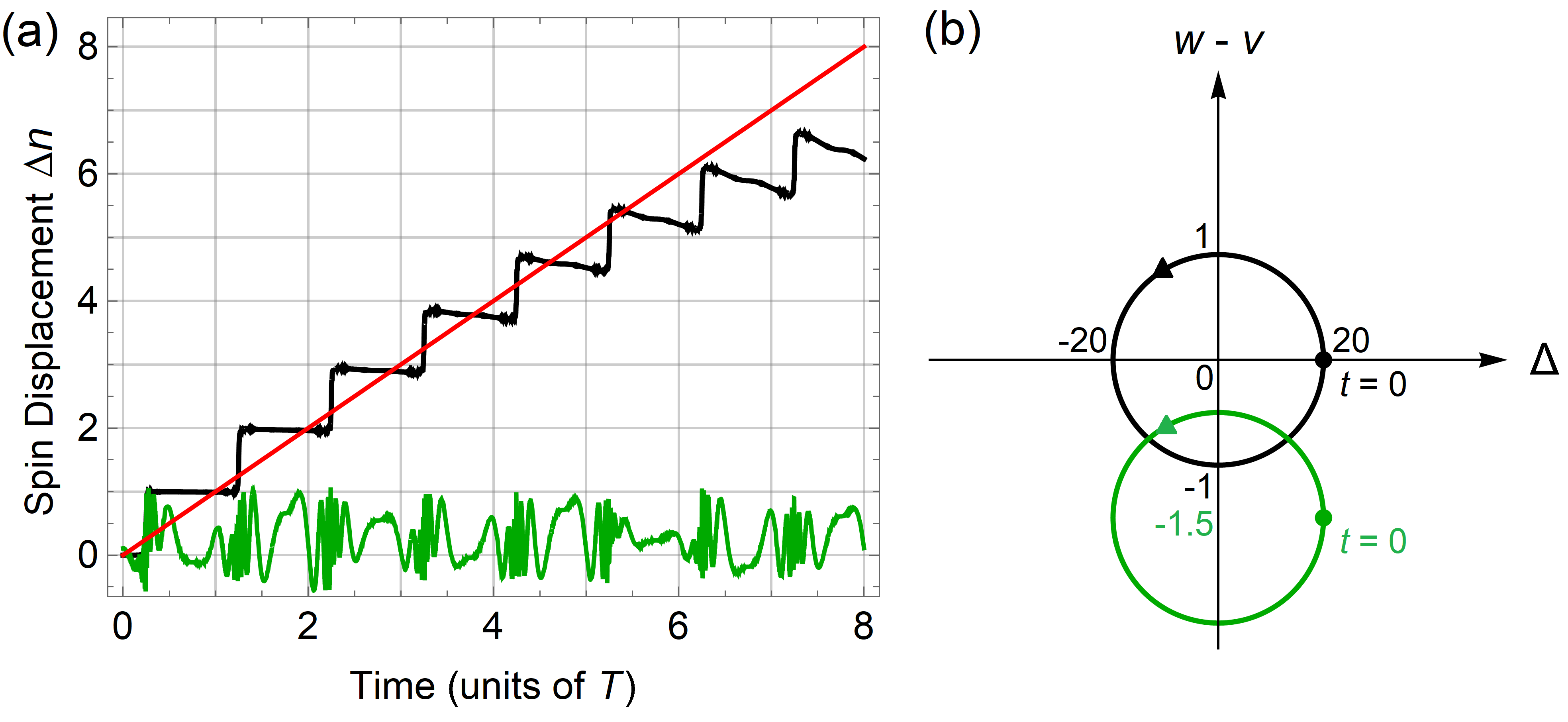}
\caption{Dynamics of spin displacement and circuit paths in parameter space. (a) Spin ($S = 50$) displacement as a function of time for a circuit in parameter space that encloses the topological transition point at $(w-v,\Delta)=(0,0)$ (black) and one that does not (green).  For the circuit that does enclose the transition point, the spin dynamics looks like a staircase as the average spin increases by one quantum of spin every cycle.  The red line indicates the slope of the ideal spin pump staircase.  The deviation from the ideal case comes from the inhomogeneity of the spin-$S$ transitions and results in a larger energy requirement to make transitions to higher spin states. (b) Paths taken through parameter space in each cycle.  The numerical values are in units of $w_0$.}
\label{fig:disp}
\end{figure}

With the addition of a spin-1/2 offset the Hamiltonian becomes 

\begin{equation}
\hat{H}_2 = \hat{H}_1 - S \Delta \hat{\sigma}_z \, .
\label{eq:offset}
\end{equation}
The new term breaks the chiral symmetry and allows for a continuous transition between the two topological phases.  The Hamiltonian resembles the Rice-Mele (RM) model which is a well known example of a Thouless charge pump.  When the RM model is adiabatically driven in a cycle of period $T$ the charges in the lattice end up being displaced by a single unit cell if the path in parameter space encloses the topological transition point at $(w-v,\Delta) = (0,0)$.  Therefore, we should expect the giant spin to change by a single quantum of spin if an adiabatic circuit in parameter space encompasses the same point.  The time dependent parameters which are used to trace the circuit are

\begin{eqnarray}
v(t) &=& v_0\left [1 -\sin (2 \pi t/T) \right ]/2, \nonumber \\
w(t) &=& w_0\left [1+\sin (2 \pi t/T) \right]/2, \nonumber \\
\Delta (t) &=& \Delta_0 \cos (2 \pi t/T)
\label{eq:cycle}
\end{eqnarray}
where $T \gg 1$ is the driving period and is large so the driving is adiabatic.  The other parameter values are $v_0=w_0$ and $\Delta_0 = 20 w_0$.  The evolution over one cycle can be imagined with the following major steps: (1) $t=0$ the system is initialized in the ground state with the spin-1/2 particle being in the state $\vert \uparrow \rangle$ due to a large offset from an applied magnetic field, (2) $0 < t \leq T/2$ interactions are ramped up and spin exchange takes place where the giant spin gains a quantum of spin and $\vert \uparrow \rangle \to \vert \downarrow \rangle$, (3) $T/2 < t \leq T$ interactions are ramped down and the applied magnetic field resets the spin-1/2 particle $\vert \downarrow \rangle \to \vert \uparrow \rangle$.  The large offset ensures the state is nearly completely locked into one of the spin-1/2 states and transitions only take place at the times $t \approx T/4$ (spin-$S$) and $t \approx 3T/4$ (spin-1/2).  The net result is that the giant spin gains one quantum of spin much the same way a charge is displaced one unit cell in the RM model.  To show the dynamics we calculate the giant spin displacement 

\begin{equation}
\Delta n(t) = \langle \psi (t) \vert \hat{S}_z \vert \psi(t) \rangle - n_0
\label{eq:SS}
\end{equation}
where $n_0 = \langle \psi(0) \vert \hat{S}_z \vert \psi(0) \rangle$ is the initial average spin.  When $S \gg 1$ the continuum approximation can be made and Eq.\ \eqref{eq:SS} can be put into another form  (appendix A)

\begin{eqnarray}
\Delta s(t) =\int_{0}^t dt^\prime j(t^\prime)
\label{eq:SS2}
\end{eqnarray}
where $\Delta s(t) = \Delta n(t)/S$ is the continuous spin displacement variable and $j(t) = \int_{-1}^{1} ds j(s,t)$ is the total probability current.  The RHS of Eq.\ \eqref{eq:SS2} is also the expression for the transported charge, $Q(t)$, of a material when $j(t)$ is the electric current averaged over the material, so $\Delta n(t)$ can be thought of as the scaled transported spin $\Delta n(t) = SQ(t)$.  This is an important connection because in topologically nontrivial systems, the total charge transported during one adiabatic cycle is proportional to the change in the winding number $Q(T) \propto \nu(T) - \nu(0)$ \cite{xiao10}, so this is where the topological nature of the pump comes from.

Equation \eqref{eq:SS} is plotted in Fig.\ \ref{fig:disp}(a) for the cycle outlined in Eq.\ \eqref{eq:cycle} (black) and for the same cycle except $v(t) =  v_0\left [4 -\sin (2 \pi t/T) \right ]/2$ (green).  A clear difference is displayed where the spin displacement increases in integer steps each period of the cycle for the former circuit and does not go beyond unity for the latter circuit.  The two circuits through parameter space are shown in Fig.\ \ref{fig:disp}(b) where the black one encloses the topologically significant point at the origin $(w-v,\Delta) = (0,0)$ and the green one does not.  Other circuits shifted along the other three axes while not encompassing the origin have also been checked and no displacement greater than unity was observed in the same time frame.  This behavior matches the quantized displacement found in the RM model and other systems which are able to act as charge pumps.  The red line is a guide for the ideal case where the black data should intercept it at integer values of the period $T$.  At early times the spin pump does follow the ideal case, however, at $t \approx 5 T$ there starts to be a large discrepancy.  The cause of this is the square root factors in Eq.\ \eqref{eq:SR} because they increase the energy required to transition to larger spin-$S$ states the farther away the state is from $n = 0$.  For states around $n = 0$, the factors act as a harmonic trap since to leading order in spin number they are $-\sqrt{N^2/4- n^2} \approx -N/2 + n^2/N$.  Similar saturation has been observed in experiments simulating the RM model with ultracold atoms in an optical lattice \cite{nakajima16} because an overall harmonic trap is required to hold the atoms in place.  The saturation was attributed to the variation of the harmonic trap becoming comparable to the bandgap.  The major difference between the experiment and the spin system is that the tunneling and the harmonic confinement in the experiment is controlled separately with different lasers whereas that is impossible here.

\subsection{Many-body system}

\begin{figure}[t]
\centering
\includegraphics[scale=0.6]{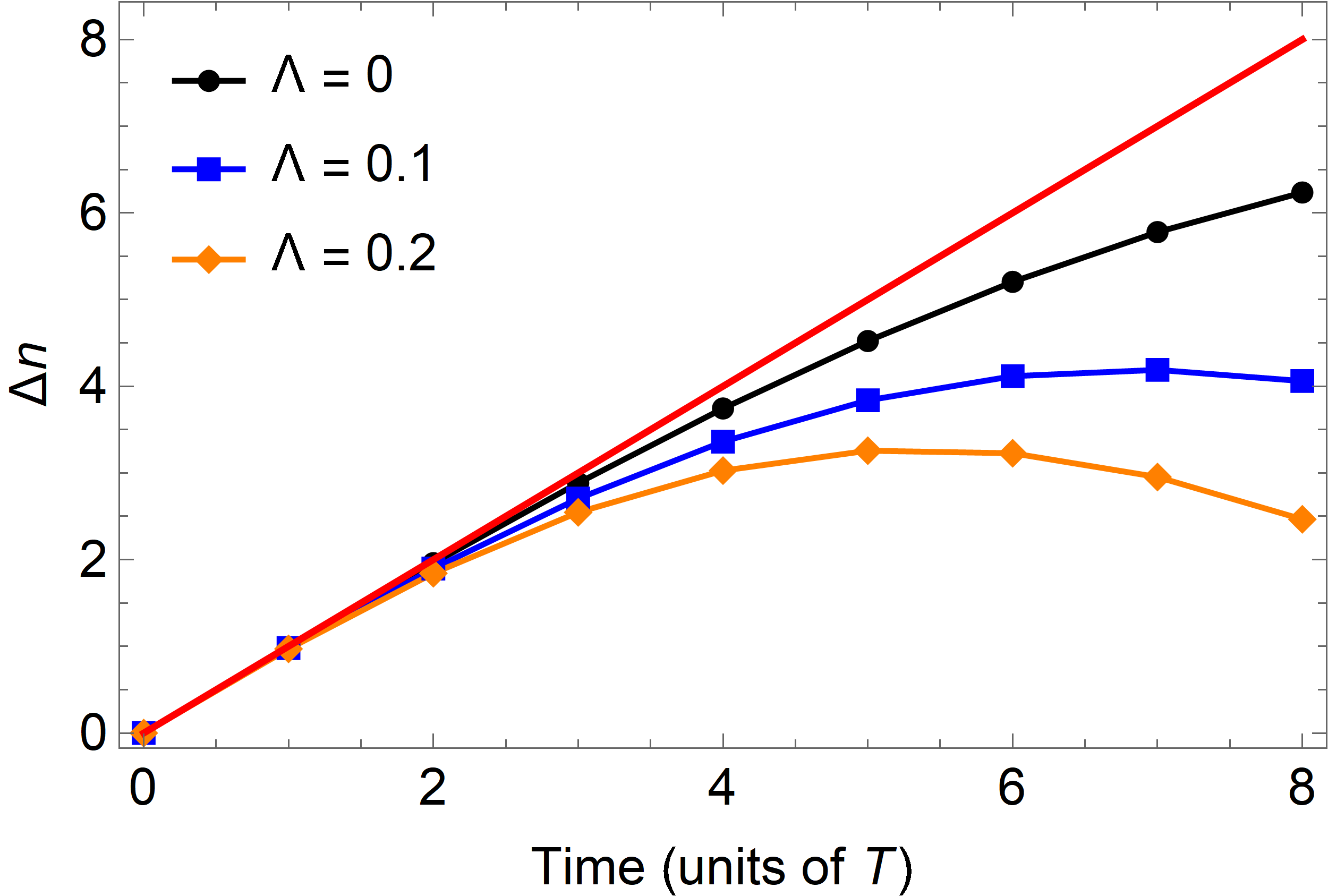}
\caption{Time dependence of the displacement for different values of the interaction energy $\Lambda$.  Each data point is the displacement at the beginning of each cycle for the path enclosing the origin in Fig.\ \ref{fig:disp} and the lines connecting the points are there as a guide.  The deviation of the displacement from the ideal case (straight red line) increases with $\Lambda$.  The number of particles is $N=2S = 100$ and the values of $\Lambda$ are in units of $w_0$.}
\label{fig:IP}
\end{figure}

The spin-$S$ operators can also be considered as many-body operators for $N = 2S$ identical spin-1/2 particles $\hat{S}_\alpha = \sum_{i=1}^{N} \hat{\sigma}_\alpha^i$ where $\alpha = (x,y,z)$.  The Pauli matrices in the sum are distinguished from the spin-1/2 ones via the superscript.  The spin-1/2 particle can represent an impurity.   In this system, the spin number becomes $n = (N_+ - N_-)/2$ which is half the difference between the number of particles in each of the two spin states labeled $+$ and $-$.  Each pump cycle is marked by one of the $N$ particles making a transition from one state to the other depending on the direction of the pump, so after one period $n \to n\pm1$.  The interaction term in Eq.\ \eqref{eq:ham} can be accomplished if each of the $N$ particles interacts with the impurity with the same strength which is commonly found in systems with infinite range interactions (interactions that do not depend on the position of each particle). We assume the pairwise interactions of the $N$ particles are infinite range as well and take the same form as their pairwise interaction with the impurity giving $-\Lambda \left (\hat{S}_+ \hat{S}_- + \hat{S}_- \hat{S}_+ \right )/N$ which can also be written as  $2\Lambda \hat{S}_z^2/N$ when $N$ is conserved.  Such interactions can be found in the infinite range isotropic XY model where the effect for $\Lambda > 0$ is harmonic confinement since there is a quadratic energy cost for states $n \neq 0$.  The additional confinement only adds to the effect of the square root factors causing the pump to substantially deviate from the ideal case at earlier times which is shown in Fig.\ \ref{fig:IP} for different values of $\Lambda$ using the circuit in Eq.\ \eqref{eq:cycle}.  To reduce clutter in the graph we only plot the displacement at the start of each cycle.

Attractive interactions ($\Lambda < 0$) energetically promote the spreading of states away from $n = 0$.  For weak attractive interactions, the ground state at $t = 0$ spreads, but still remains centered near $n = 0$ due to the square root factors dominating.  However, there is a critical interaction strength where a quantum phase transition (QPT) occurs resulting in the shift of the center of the ground state to $n \neq 0$.  The critical value is (Appendix B)

\begin{equation}
\Lambda_c = - \frac{w (w+v)}{\sqrt{\Delta^2 + (w + v)^2}} \, .
\label{eq:Lcrit}
\end{equation}

Care must be taken when $\Delta \gg w, v$, which is what we have used, because the system is sensitive to any attractive interactions since the critical value $\Lambda_c \approx w(w+v)/\Delta$ will be close to zero.  Preparing the initial state under these conditions could result in a large shift of the center of the state toward the edges at $n = \pm N/2$ where the winding number drops rapidly, as seen in the bottom panel of Fig.\ \ref{fig:BP}(c), resulting in the destruction of the pump.  This issue can be resolved by simply preparing the initial state in another segment of the cycle around $\Delta \approx 0$.  However, if $\Lambda < \Lambda_c^\mathrm{min}$, where $\vert \Lambda_c^\mathrm{min} \vert$ is the minimum interaction strength over one cycle, then the system will ramp through the critical point twice per cycle resulting in unwanted excitations.

\section{Conclusion}

We have shown that a system consisting of spinful particles interacting via spin exchange can simulate inhomogeneous versions of the SSH and RM models.  The spin system shares all of the topological features of both models which includes topologically protected edge states and a quantized spin pump in the form of a Thouless pump.  We also showed that when the spin system is interpreted as a many-body system, repulsive interactions lead to decrease efficiency of the pump.    The majority of simulations of condensed matter systems depend on spatial degrees of freedom such as optical lattices and resonator arrays to simulate the crystal structure of solids, however, all that is needed are discrete sets of states which are abundant in quantum systems.   Here, we have presented an extreme case of this by absorbing all of the spatial dependence into the internal spin states of two particles.

The large values of $S$ throughout this paper are used to highlight the topological features of the model.  In reality, the spins of individual particles are much smaller, so the topological transition point at $(w-v,\Delta)=(0,0)$ becomes a transition region around this point due to finite size effects.  This means the smaller $S$ is, the smaller $v$ must be to observe the edge states and the larger the circuit must be to observe the spin pump.  Of course, a bulk must exist which requires $S>1$ since there should be two bulk unit cells to pump between.  Therefore, the minimum case has $n = \pm 3/2$ and $n = \pm 1/2$ as the edge and bulk  states, respectively.  To observe larger $S$ effects, the infinite range XY model can be used where the spin-1/2 particle plays the role of an impurity.  We found the best spin pump result is achieved when there are no interactions between the $N$ identical particles in the XY chain, but there are equal interactions between the particles in the chain and the impurity.  This can happen in a variation of the central spin model \cite{hassanieh06} where the chain makes a horseshoe shape with the impurity at the center, so the impurity is roughly equal distance from each particle along the chain and the edges are maintained.    
\appendix

\section{Derivation of $\Delta s(t)$}
We start with Eq.\ \eqref{eq:SS} and expand the $\vert \psi \rangle$ states in terms of the $\hat{S}_z$ eigenstates, $\vert \psi(t) \rangle = \sum_n c_n(t) \vert n \rangle$ where $c_n(t) = \langle n \vert \psi(t) \rangle$, giving

\begin{equation}
\Delta n (t) = \sum_n n \left [\vert c_n(t) \vert^2  - \vert c_n (0) \vert^2 \right ] \, .
\label{eq:ss1}
\end{equation}
The continuum approximation is used by assuming $S \gg 1$, so that $s=n/S$ is a continuous variable.  We also identify $\rho (s,t) = \vert c_n(t) \vert^2$ as the probability density where $\int_{-1}^1 ds \rho(s,t) = 1$, so Eq.\ \eqref{eq:ss1} becomes 

\begin{equation}
\Delta s (t) =\int_{-1}^1 ds s \left [ \rho(s,t) - \rho(s,0) \right ] \, .
\label{eq:ss2}
\end{equation}
The probability density difference can be put into integral form $\rho(s,t) - \rho(s,0) = \int_{0}^{t} dt^\prime \partial \rho(s,t^\prime)/\partial t^\prime$ and the probability continuity equation, $\partial \rho(s,t)/\partial t = -\nabla j(s,t)$, is used to give

\begin{equation}
\Delta s (t) = - \int_0^t dt^\prime \int_{-1}^1 ds s \nabla j(s,t^\prime) \, . 
\label{eq:ss3}
\end{equation}
Next, we use integration by parts and set the boundary terms to zero by making the assumption that the probability current falls off at the edges of the system which leads to the final result

\begin{equation}
\Delta s (t) =  \int_0^t dt^\prime \int_{-1}^1 ds j(s,t^\prime) \, . 
\label{eq:ss3}
\end{equation}
Defining $j(t) = \int_{-1}^1 ds j(s,t)$ as the total probability current gives Eq.\ \eqref{eq:SS} in the main text.

\section{Derivation of $\Lambda_c$}

\begin{figure}[t]
\centering
\includegraphics[scale=0.6]{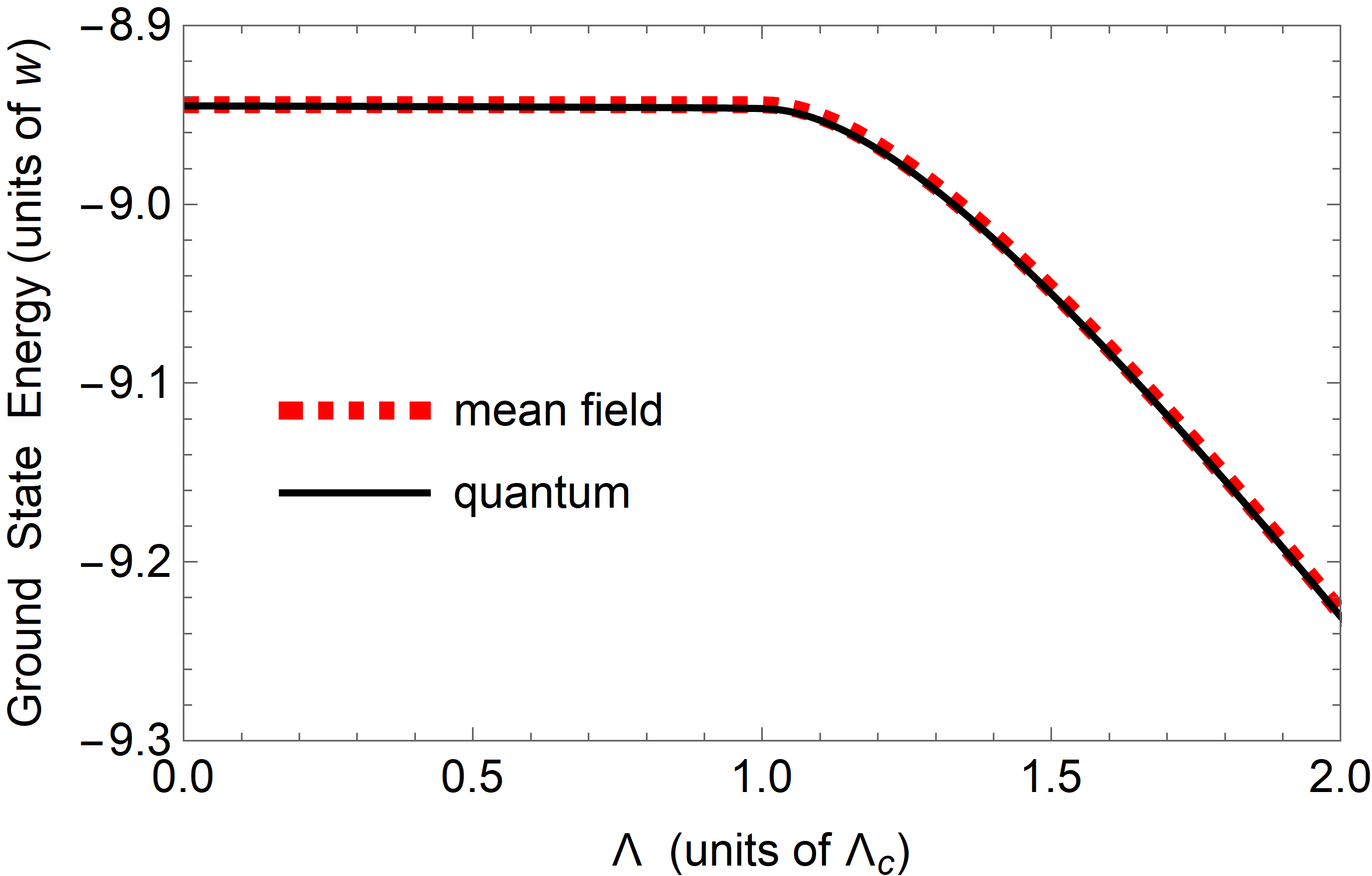}
\caption{Quantum and mean field ground state energies.  The ground state energy shows a shift around the critical interaction energy of $\Lambda_c$ signaling a QPT.  The parameters used in the calculations are $v=1$, $\Delta=4$ (both in units of $w$) and $N = 400$. }
\label{fig:QPT}
\end{figure}

We are interested in the ground state QPT, so we start by finding an expression for the ground band energy.  The Hamiltonian including interactions is

\begin{equation}
\hat{H}_3 = \frac{2 \Lambda}{N} \hat{S}_z^2 - w \left (\hat{S}_+ \hat{\sigma}_- + \hat{S}_- \hat{\sigma}_+ \right ) -N v \hat{\sigma}_x/2 -N \Delta \hat{\sigma}_z/2 
\label{eq:ham3}
\end{equation}
and making a mean field approximation by replacing the spin-$S$ operators with their spin coherent expectation values gives

\begin{equation}
\frac{ 2 H_3}{N} = \frac{\Lambda}{2}\cos\theta^2 - w \sin\theta \left ( e^{i \phi} \hat{\sigma}_- + e^{-i \phi} \hat{\sigma}_+ \right ) - v \hat{\sigma}_x - \Delta \hat{\sigma}_z 
\end{equation}
where we have written the spin number in terms of the polar angle on the Bloch sphere $n = (N/2) \cos\theta$.  Setting $\phi = 0$ minimizes the Hamiltonian further and diagonalizing it gives an expression for the ground band energy

\begin{equation}
\frac{2 E_0(\theta)}{N} = \frac{\Lambda}{2} \cos\theta^2 -  \sqrt{\Delta^2+(v+w\sin\theta)^2} \, .
\label{eq:GBE}
\end{equation}
It is expected that the phase transition occurs when there is a new ground state.  Due to the symmetry around the ground state at  $\theta = \pi/2$ ($n = 0$), we expect a second order phase transition in the ground band where the initial ground state transforms into a local maximum and two new degenerate $\theta \neq \pi/2$ ground states appear when $\Lambda < \Lambda_c$.  We expand $E_0(\theta)$ around $\theta = \pi/2$

\begin{equation}
\frac{2 E_0(\theta)}{N}\approx A + B \delta \theta^2 + C \delta\theta^4 
\end{equation}
where $\delta\theta = \theta - \pi/2$ and note that new minima occur when $B<0$ while $C$ is positive.  The constant $B$ is

\begin{equation}
B = \Lambda+ \frac{w (w+v)}{\sqrt{\Delta^2 + (w + v)^2}} \, ,
\end{equation}
so $\Lambda_c$ is Eq.\ \eqref{eq:Lcrit} in the main text.  A phase transition is marked by a shift in the ground state energy which we plot in Fig.\ \ref{fig:QPT} where the black solid curve is calculated by diagonalizing Eq.\ \eqref{eq:ham3} and the dashed red curve is calculated by finding the minimum of Eq.\ \eqref{eq:GBE}.  The image shows excellent agreement between the quantum and mean field predictions and that there is a shift in the ground state energy at $\Lambda \approx \Lambda_c$.

\end{document}